\documentstyle[sprocl,psfig,epsfig]{article}

\def\beq{\begin{equation}}
\def\eeq{\end{equation}}
\def\bea{\begin{eqnarray}}
\def\eea{\end{eqnarray}}

\begin{document}

\title{HADRON CORRELATORS IN THE DECONFINED PHASE\footnote{Talk given at the 
``Strong and Electro-Weak Matter'' Conference (14-17 June 2000, Marseille)}}

\author{F. KARSCH}

\address{Fakult\"at f\"ur Physik, Universit\"at Bielefeld,
D-33615 Bielefeld, Germany\\E-mail: karsch@physik.uni-bielefeld.de} 

\author{M.G. MUSTAFA}

\address{Theoretical Nuclear Physics Division, Saha Institute of Nuclear 
Physics, 1/AF Bidhan Nagar, Calcutta - 700 064, India\\E-mail: 
mustafa@tnp.saha.ernet.in}

\author{M.H. THOMA}

\address{Theory Division, CERN, CH-1211 Geneva 23, Switzerland\\E-mail: 
markus.thoma@cern.ch}

\maketitle\abstracts{Temporal meson correlators and
their spectral functions are calculated in the deconfined phase
using the hard thermal loop  resummation technique. The spectral 
functions exhibit strong medium effects coming from the hard thermal loop 
approximation for the quark propagator. The correlators, on the other hand,
do not differ significantly from free correlators, for which
bare quark propagators are used. This is in contrast to lattice
calculations showing a clear deviation from the free correlations functions.}


\section{Introduction}
QCD Lattice calculations have been used successfully for studying hadron properties at
zero temperature. At finite temperature there is a sudden change
of these properties as soon as the critical temperature $T_c$ for the deconfinement
transition has been reached. However, in particular in the pseudo-scalar channel
the mesonic correlation functions differ clearly from free correlators, describing freely 
propagating bare quarks, even above $T_c$ \cite{Boy94}. These deviations could be caused
either by bound states of quarks above $T_c$ or by in-medium modifications of the collective quark
modes in the QGP. In order to investigate the origin of the non-trivial correlations
we calculate temporal meson correlations functions 
using the hard thermal loop (HTL) resummation technique \cite{Bra90}. This improved perturbation theory
for QCD at high temperatures ($T\gg T_c$) takes into account important medium effects of the
QGP such as effective, temperature dependent quark masses and Landau damping \cite{LeB96}. 
Therefore it is interesting to investigate to what extent the HTL approximation for the mesonic 
correlators is able to explain the lattice results. 
In particular temporal correlators and their spectral 
functions, which we will consider in the following exclusively, yield interesting information about 
hadronic properties at finite temperature.

\vspace*{-0.2cm}

\section{Meson Correlators and Spectral Functions}

Meson correlators are defined as expectation values 
\beq
G_M(\tau,\vec{x})= \langle J_M (\tau, \vec{x}) J_M^{\dagger} (0, \vec{0}) 
\rangle 
\label{eq1}
\eeq
of meson currents $J_M (\tau,\vec{x}) =\bar{q}(\tau, \vec{x})\Gamma_M q(\tau, \vec{x})$ 
in Euclidean time $\tau =it\,\epsilon \, [0, 1/T]$. Here $q(x)$ denotes the quark
wave function and $\Gamma_M =1$, $\gamma_5$, $\gamma_\mu$, $\gamma_\mu \gamma_5$
the quark-meson vertex corresponding to the mesonic channel (scalar, pseudo-scalar, vector,
pseudo-vector) under consideration.

The temporal correlator, defined as the limit $\vec{x}=0$ of the correlation 
function, is related to the correlation function in momentum space by 
\beq
G_M(\tau,\vec{x}=0) = T \sum_{n=-\infty}^{\infty}
\;{\rm e}^{-i\omega_n \tau}\;
\chi_M(\omega_n,\vec{p}=0),
\label{eq4}
\eeq
where $\omega_n=2\pi nT$ are the bosonic Matsubara frequencies.

An interesting quantity which we will study in the following
is the spectral function of the mesonic correlator. As we will see, 
the spectral function contains much more information about
medium effects in the QGP than the correlator itself. It is defined
as
\beq
\chi_M(\omega_n,\vec{p}) = -\int_{-\infty}^{\infty} {\rm d}
\omega\; {\sigma_M(\omega,\vec{p}) \over i\omega_n - \omega +i\epsilon}.
\label{eq5}
\eeq
Using (\ref{eq4}) and (\ref{eq5}) the temporal correlator can be expressed
by
\beq
G_M(\tau) =  \int_{0}^{\infty} {\rm d} \omega\; 
\sigma_M (\omega,\vec{p}=0)\; 
{{\rm cosh}(\omega (\tau - 1/2T)) \over {\rm sinh} (\omega /2T)}.
\label{eq7}
\eeq


\section{Free Correlators}

\begin{figure}
\label{fig:bubble}
\begin{center}
~\epsfig{file=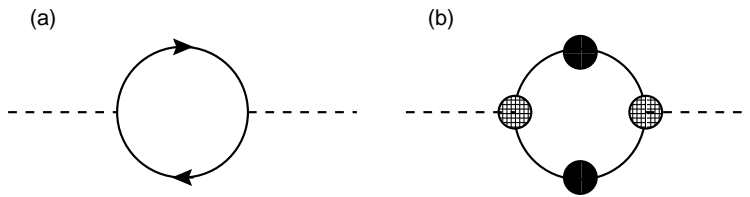,width=100mm}
\end{center}
\caption{Meson correlators for free quarks (a) and in the
HTL approximation (b).}
\end{figure}
The free mesonic correlators follow from the one-loop self-energy
diagram of Fig.1(a) containing bare quark propagators. In momentum space 
the correlation function reads
\beq
\chi_M(\omega, \vec{p}) = 2 N_c 
T \sum_n \int {{\rm d}^3k \over (2 \pi)^3} \;{\rm Tr}\biggl[ \Gamma_M
S_F(k_0,\vec{k}) \Gamma_M^{\dagger} 
S_F^{\dagger} (\omega - k_0, \vec{p}-\vec{k}) \biggr ],
\label{eq8}
\eeq
where $k_0=i(2n+1)\pi T$ are the fermionic Matsubara frequencies.
This leads to the following expression for the spectral functions of the free 
temporal correlator 
\beq
\sigma_M^{\rm free} (\omega) = 
{N_c \over 4 \pi^2} \> \theta (\omega- 2m) \>  \omega^2  \>\tanh \left 
(\frac{\omega}{4T}\right)\> 
\sqrt{1-\biggl({2m\over \omega}\biggr)^2 }
\> \biggl[a_M + \biggl({2m \over \omega}\biggr)^2 b_M\biggr], 
\label{eq9}
\eeq
where $\sigma_v={\sigma_v}^\mu_\mu$ in the (pseudo-)vector case.
The coefficients $(a_M,b_M)$ are given by $(-1,1)$, $(1,0)$, $(2,1)$, 
and $(-2,3)$ in the scalar, pseudo-scalar, vector, and pseudo-vector case,
respectively. For massless bare quarks, $m=0$, $|\sigma_{ps}| = |\sigma_{s}|$ 
and $|\sigma_{pv}| = |\sigma_{v}|$ hold, reflecting chiral symmetry.
For $m=0$ analytic expressions can be derived \cite{Flo94}
for the temporal correlator
from (\ref{eq9}) using (\ref{eq7}). For large
real times $t \rightarrow \infty$ they reduce to
$G_M(t) \sim \exp (-\mu t)$, with the screening mass
$\mu =2\pi T$. The same screening mass has also been found
for spatial correlators at large distances \cite{Flo94}. 


\section{HTL Approximation}

Now we want to consider the HTL approximation for the temporal correlator
and its spectral function using HTL resummed propagators and vertices as
shown in Fig.1(b). In this way important medium effects of the high temperature plasma, 
namely effective temperature dependent quark masses and Landau damping are taken
into account. The HTL quark propagator in the imaginary time formalism
can be written as
\beq
S_F^{\rm HTL}(k_0, \vec{k})=-(\gamma_0 \, k_0  - \vec{\gamma}\cdot \vec{k})
\int_0^{1/T}\! \! {\rm d}\tau \> {\rm e}^{k_0\tau} 
\int_{-\infty}^{\infty} \! \! {{\rm d}\omega} \>
\rho_{\rm HTL} (\omega, \vec{k})\> [1-n_F(\omega)]\> {\rm e}^{-\omega \tau},
\label{eq11}
\eeq
where $n_F(\omega) = 1/(1+\exp(\omega / T))$ is 
the Fermi distribution. It is convenient
to decompose the spectral function of the HTL quark propagator according to
its helicity eigenstates \cite{Bra90a}
\beq
\rho_{\rm HTL} (\omega, \vec{k}) =  
{1\over 2} \rho_{+} (\omega, k)(\gamma_0 - i\; \hat{k}\cdot \vec{\gamma}) 
+{1\over 2} \rho_{-} (\omega,k)(\gamma_0 + i\; \hat{k}\cdot \vec{\gamma}),
\label{eq12}
\eeq
where
\beq
\rho_{\pm} (\omega, k) = {\omega^2 - k^2 \over 2 m_q^2} 
[\delta (\omega - \omega_{\pm}) + \delta (\omega + \omega_{\mp})]
+\beta_{\pm} (\omega, k) \theta(k^2 -\omega^2).
\label{eq13}
\eeq
Here the first term of (\ref{eq13}) 
comes from the poles of the HTL propagator with
the in-medium quark dispersion relation $\omega_\pm(k)$. This dispersion
relation exhibits two branches, of which the upper one ($\omega_+$)
corresponds to a collective quark mode and the lower one ($\omega_-$)
to a plasmino having a 
negative ratio of helicity to chirality and being absent in vacuum
\cite{Bra90a}. The plasmino dispersion shows a minimum at non-vanishing
momentum as a general property of massless fermions in relativistic
plasmas \cite{Pes00}. The cut
contribution, coming from the imaginary part of the quark self energy
which describes Landau damping, reads
\beq
\beta_{\pm} (\omega, k) = -{m_q^2 \over 2}\> { \pm \omega - k\over
\biggl[k(-\omega \pm k) + m_q^2 \biggl( \pm 1 - {\pm \omega -k \over 2k}
\ln{k+\omega \over k-\omega} \biggr)\biggr]^2 + \biggl[ {\pi \over 2}
m_q^2 {\pm \omega -k \over k} \biggr]^2 },
\label{eq14}
\eeq
where the effective quark mass in the HTL 
approximation is given by $m_q=g(T) T/\sqrt{6}$. 

According to the above decomposition of
the quark propagator in a pole and a cut contribution, the spectral
function of the temporal correlator following from Fig.1(b) can be
written as a sum of a pole-pole, pole-cut, and a cut-cut term,
$\sigma^{\rm HTL}  (\omega) = \sigma^{\rm pp} (\omega) +  
\sigma^{\rm pc} (\omega) +  \sigma^{\rm cc} (\omega)$.
The most interesting term is the pole-pole contribution,
for which we obtain for the pseudo-scalar spectral function
\bea
&&\sigma^{\rm pp}_{\rm ps} (\omega)=
{N_c\over 2\pi^2 m_q^4} ({\rm e}^{\omega/T} -1)\> 
\biggl[ n_F^2(\omega_+(k_1)){(\omega_+^2(k_1) -k_1^2)^2\, 
k_1^2\over 2| \omega_+'(k_1)|}
\nonumber \\
&&+2 \sum_{i=1}^2
n_F(\omega_+(k_2^i))[1-n_F(\omega_-(k_2^i))] 
{(\omega_-^2(k_2^i) -(k_2^i)^2)\, (\omega
_+^2(k_2^i) -(k_2^i)^2)\,
(k_2^i)^2\over | \omega_+'(k_2^i)-\omega_-'(k_2^i)|}
\nonumber \\
&&+\sum_{i=1}^2
n_F^2(\omega_-(k_3^i)) {(\omega_-^2(k_3^i) -(k_3^i)^2)^2\, 
(k_3^i)^2\over 2| \omega_-'(k_3^i)|}\biggr].
\label{eq15}
\eea
Here we have neglected the HTL vertex correction in Fig.1(b)
since in the case of the (pseudo-)scalar correlator it leads to higher 
order corrections only, as in Yukawa theory \cite{Tho95}. The momenta $k_n^i$
are determined from the zeros of the equations
$\omega - 2 \omega_+(k_1) = 0$,
$\omega - \omega_+(k_2^i)+\omega_-(k_2^i) =0$, and  
$\omega - 2  \omega_-(k_3^i) = 0$.
Due to the minimum in the plasmino dispersion two solutions for $k_2^i$ 
and $k_3^i$ ($i=1,2$) are possible. 
The first term in (\ref{eq15}) describes the annihilation of collective quarks,
the second one the transition from the upper to the lower branch, and
the last one the annihilation of plasminos. The transition process starts at
zero energy and continues until the maximum difference $\omega =0.47\> m_q$
between the two branches at $k_2=1.18\> m_q$. At this point a Van Hove 
singularity is encountered due to the vanishing denominator
$\omega'_+(k_2)-\omega'_-(k_2) =0$, where the prime denotes the derivative
with respect to $k$. The plasmino annihilation starts at $\omega = 1.86\> m_q$
with another Van Hove singularity corresponding to the minimum of the plasmino
branch at $k_3=0.41\> m_q$, where $\omega'_-(k_3) = 0$. This contribution falls off rapidly due to the 
exponentially suppressed spectral strength of the plasmino mode for large energies,
where only the first process, quark-antiquark annihilation starting at 
$\omega =2m_q$, contributes.
The pole-cut and cut-cut contributions, on the other hand, are smooth 
functions of $\omega$. The individual contributions to the spectral function of the temporal 
pion correlator are shown in the left panel of Fig.2 for $m_q/T=1$. 

\begin{figure}
\label{fig:sigma}
\begin{center}
\epsfig{file=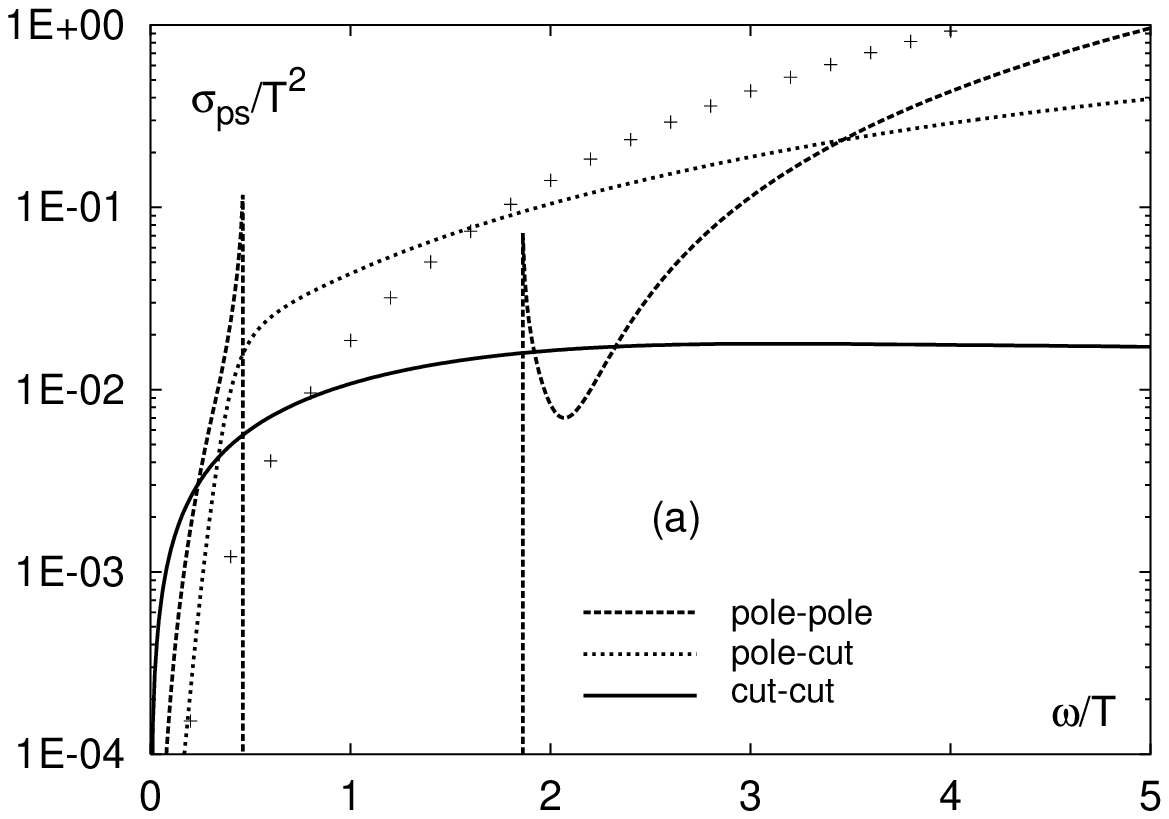,width=59mm}
\epsfig{file=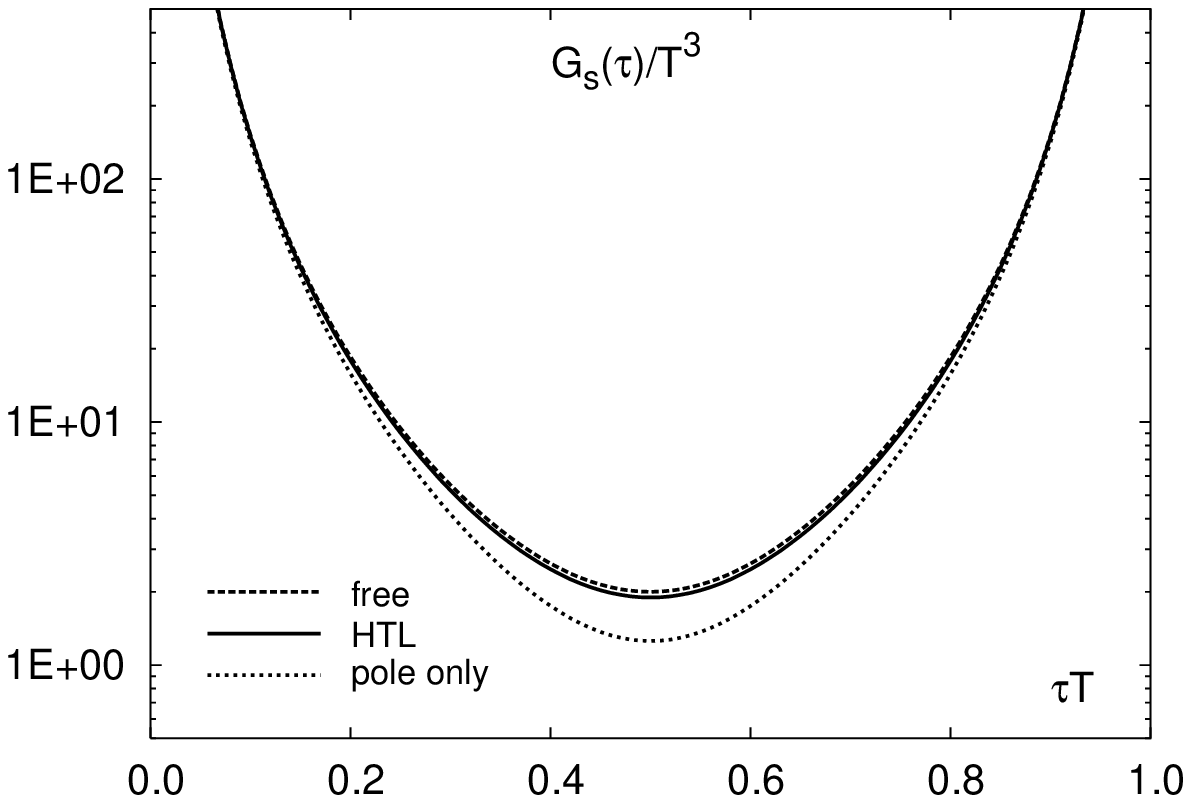,width=59
mm}
\end{center}
\caption{Left panel: The pole-pole, pole-cut and cut-cut contributions to the 
pseudo-scalar spectral function for $m_q/T=1$. 
The crosses show the free meson spectral function.
Right panel: The thermal pseudo-scalar meson correlation function in the HTL 
approximation for $m_q/T=1$. The curves show the 
complete thermal correlator (middle line), the correlator constructed from 
$\sigma^{\rm pp}_{\rm s}$ only (lower line) and the free thermal correlator 
(upper line).} 
\end{figure} 

Surprisingly the temporal correlator following from integrating over $\omega $ according
to (\ref{eq7}) is similar to the free one in spite of the significant structures
showing up in the spectral function (right panel of Fig.2). There is a cancellation of the
(due to the effective quark mass) reduced pole-pole contribution compared to the
free correlator and the additional pole-cut and cut-cut contributions
to a large extent. 

A similar behavior can be observed for
the temporal vector correlator \cite{Kar00}. In this
case a HTL quark-meson vertex has to be considered as in the case of the closely
related soft dilepton production rate \cite{Bra90a}. The pole-pole contribution
to the spectral function contains again Van Hove singularities at the same 
energies as in the pseudo-scalar case. However, the cut-cut contribution
diverges now for small energies like ${\sigma_v}^{\rm cc}(\omega ) 
\sim 1/\omega$ leading to a singular result for the temporal vector 
correlator after integrating over $\omega $. 


\section{Conclusions}

Lattice calculations of meson correlators in the deconfined phase show
a clear deviation from the free correlation functions in particular
in the pseudo-scalar channel. The long range behavior of free correlators 
in time and space can be described by a screening mass $\mu = 2\pi T$
corresponding to the propagation of two bare quarks. Using HTL 
resummed quark propagators instead of bare ones, medium effects, 
such as thermal quark masses and Landau damping, due to 
interactions with the thermal particles of the QGP are taken into account.
This leads to sharp structures (Van Hove singularities, energy gap)
in the spectral functions of the temporal correlators. 
However, the correlators following from their spectral functions
by an energy integration are close to the free correlators due to a
partial cancellation of the pole-pole and the cut contributions
in the HTL approximation. In other words, HTL medium effects appear not 
to be sufficient to explain the lattice results. This supports 
speculations about bound states or other non-perturbative effects
in the QGP close to the critical temperature.

\section*{Acknowledgments}
FK has been supported by the TMR network ERBFMRX-CT-970122 and the DFG
under grant Ka 1198/4-1. MGM would like to acknowledge support from AvH
Foundation as part of this work was initiated during his stay at the University
of Giessen as Humboldt Fellow. MHT has been supportd by the DFG as Heisenberg
Fellow.


\section*{References}
\begin{thebibliography}{99}

\bibitem{Boy94} G. Boyd, S. Gupta, F. Karsch and E. Laermann,
{\em Z. Phys.} C {\bf 64}, 331 (1994).
\bibitem{Bra90} E. Braaten and R.D. Pisarski, {\em Nucl. Phys.} {\bf B} 337, 569 (1990).
\bibitem{LeB96} M. Le Bellac, {\em Thermal Field Theory} (Cambridge
University Press, Cambridge, 1996).
\bibitem{Flo94} W. Florkowski and B.L. Friman, {\em Z. Phys.} A {\bf 347}, 271 (1994).
\bibitem{Bra90a} E. Braaten, R.D. Pisarski and T. C. Yuan, 
{\em Phys. Rev. Lett.} {\bf 64}, 2242 (1990).
\bibitem{Pes00} A. Peshier and M.H. Thoma, {\em Phys. Rev. Lett.} {\bf 84}, 841 (2000).
\bibitem{Tho95} M.H. Thoma, {\em Z. Phys. C} {\bf 66}, 491 (1995).
\bibitem{Kar00} F. Karsch, M.G. Mustafa, and M.H.Thoma, {\em hep-ph/0007093}.


\end{thebibliography}
\end{document}